\documentclass[conference]{IEEEtran}
\usepackage{comment}

\usepackage{cite}
\usepackage{amsmath,amssymb,amsfonts}
\usepackage{algorithmic}
\usepackage{graphicx}
\usepackage{textcomp}

\usepackage{cite}
\usepackage{amsmath,amssymb,amsfonts}
\usepackage{algorithmic}
\usepackage{graphicx}
\usepackage{epstopdf}
\usepackage{subfigure}
 \graphicspath{{\figures}}
\usepackage{textcomp}
\usepackage{xcolor}
\usepackage{verbatim}
\usepackage{makecell}
\usepackage{booktabs}
\usepackage{CJK}
\usepackage{multirow}

\usepackage{algorithm}
\usepackage{algorithmic}

\makeatletter
\newcommand{\removelatexerror}{\let\@latex@error\@gobble}
\makeatother

\IEEEoverridecommandlockouts   
\begin{document}


\title{Image Segmentation Semantic Communication over Internet of Vehicles}

\author{Qiang Pan, Haonan Tong, Jie Lv,  Tao Luo, Zhilong Zhang,\\ Changchuan Yin, and Jianfeng Li\\
\small 
Beijing Laboratory of Advanced Information Network,\\
Beijing Key Laboratory of Network System Architecture and Convergence, \\
Beijing University of Posts and Telecommunications, Beijing, China 100876.\\
Emails: \{pq569375378, hntong, lvj, tluo, ccyin, lijf\}@bupt.edu.cn, zhilong.zhang@outlook.com
\vspace{-0.4cm}
}

\maketitle
\vspace{-2cm}

\begin{abstract}
In this paper, the problem of semantic-based efficient image transmission is studied over the Internet of Vehicles~(IoV).
In the considered model, a vehicle shares massive amount of visual data perceived by its visual sensors to assist other vehicles in making driving decisions. 
However, it is hard to maintain a high reliable visual data transmission due to the limited spectrum resources.
To tackle this problem, a semantic communication approach is introduced to reduce the transmission data amount while ensuring the semantic-level accuracy. 
Particularly, an image segmentation semantic communication~(ISSC) system is proposed, which can extract the semantic features from the perceived images and transmit the features to the receiving vehicle that reconstructs the image segmentations.
The ISSC system consists of an encoder and a decoder at the transmitter and the receiver, respectively.
To accurately extract the image semantic features, the ISSC system encoder employs a Swin Transformer based multi-scale semantic feature extractor. 
Then, to resist the wireless noise and reconstruct the image segmentation, a semantic feature decoder and a reconstructor are designed at the receiver.
Simulation results show that the proposed ISSC system can reconstruct the image segmentation accurately with a high compression ratio, and can achieve  robust transmission performance against channel noise, especially at the low signal-to-noise ratio~(SNR).
In terms of mean Intersection over Union~(mIoU), the ISSC system  can achieve an increase by 75\%, compared to the baselines using traditional coding methods.
 
\end{abstract}

\section{Introduction}
 
In the emerging Internet of Vehicles~(IoV), there is a massive amount of visual information required to be transmitted among the vehicles to assist driving decisions, which consumes a certain amount of bandwidth resources\cite{7786130}, \cite{9838913}.
However, due to the restricted spectrum resources and the complicated communication conditions in the traffic environment, it is difficult to  maintain a reliable connection to transmit large amounts of visual data. 
To this end, semantic communication\cite{8869705,shi2021semantic}, with transmitting the data at a semantic level, rather than the symbol level, is becoming a viable solution to solve the above challenges.  
In current IoV system, vehicles mainly use visual data for locating and obstacle avoidance, it is available for the receiving vehicle to only reconstruct the image segmentation for making driving decisions.
According to our research, although semantic communication can reduce the data transmission amount thus reducing the occupied spectrum, the image segmentation semantic communication~(ISSC) system has not been well designed. 
 
Existing works \cite{carnap1952outline,bao2011towards,guler2018semantic,xie2021deep,tong2021federated} have systematically studied semantic communication.
In\cite{carnap1952outline}, the authors introduced a preliminary theory of semantic information based on logical probabilistic ranking. 
The work in\cite{bao2011towards} presented a model theory based technique for semantic data compression and trustworthy semantic communication with quantitative measurements.
Besides, the authors in\cite{guler2018semantic}  modeled the communication problem as a Bayesian game to minimize the end-to-end average semantic metric in a dynamic communication scenario. 
Furthermore, the work in\cite{xie2021deep} developed a deep learning based semantic communication system for text transmission, which recovered the meaning of sentences, rather than the transmitted bits or symbols. 
Recently, a semantic communication system for transmitting audio in the Internet of Things~(IoT) is proposed, and federated learning is employed to improve the precision of semantic data extraction \cite{tong2021federated}. 
However, these aforementioned works did not explore image semantic communication, and image semantic communication is challenging due to the complexity of image structure.

Along with the development of deep learning, the prior arts have investigated image semantic communication systems\cite{bourtsoulatze2019deep,kurka2020deepjscc,yang2021semantic,kang2022task}.
In\cite{bourtsoulatze2019deep}, the authors proposed an end-to-end joint source-channel coding~(JSCC) image system on the structure of autoencoder, which provided a smooth performance degradation with the decrease of signal-to-noise ratio~(SNR).
Based on JSCC, the work in\cite{kurka2020deepjscc} exploited the channel feedback for image transmission, and provided considerable improvements in terms of the end-to-end image reconstruction quality.
Besides, the work in \cite{yang2021semantic} developed a framework for semantic communication with artificial intelligence tasks and built a system for inspecting surface defects of workpieces, which realized a visual semantic communication. 
Moreover, The authors in \cite{kang2022task} extended the semantic communication for image classification tasks to UAV aerial photography and achieved the tradeoff between transmission delay and classification accuracy.
However, the works in \cite{bourtsoulatze2019deep,kurka2020deepjscc,yang2021semantic,kang2022task} were mainly concerned with how to reconstruct or classify the images accurately but did not propose a system to reconstruct the image segmentation from semantic features at the receiver directly.


\begin{figure}[t]
\centering
\includegraphics[width=8cm,height=6cm]{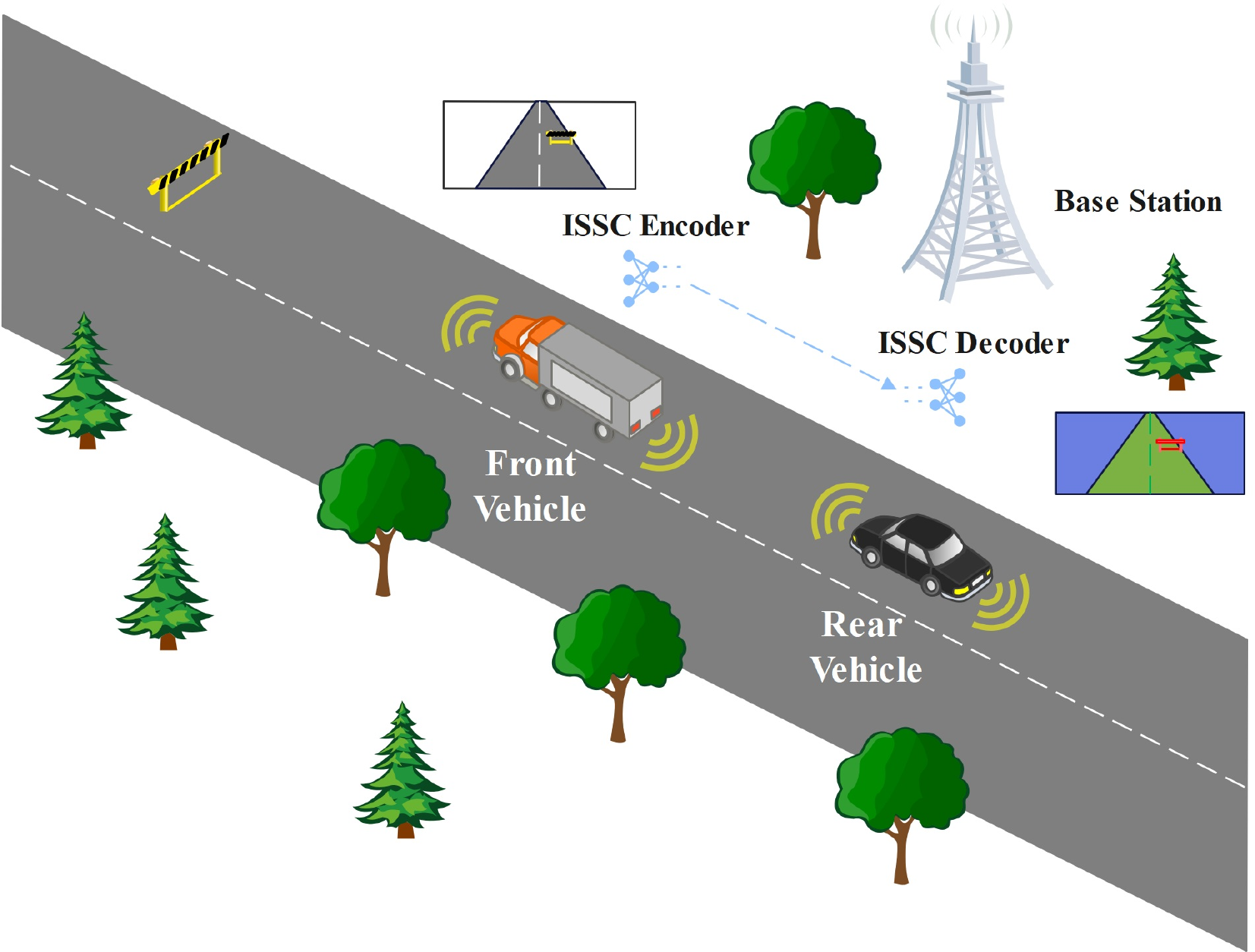}
\caption{Single-link vehicle-to-vehicle scenario with an ISSC system over the IoV.}
\label{Traffic Scene}
\vspace{-0.35cm}
\end{figure}

In this paper, a novel ISSC system is proposed for the IoV.
To our best knowledge, \textit{this is the first semantic communication system to reconstruct image segmentation.}
The main contributions are as follows:
\begin{itemize}
\item We propose an ISSC system that enables the sharing of visual data  over the IoV.
The ISSC system consists of an encoder that extracts the image semantic features at the transmitter and a decoder which reconstructs the image segmentation at the receiver. 
\item We propose a cascade structure to extract multi-scale semantic features of image based on Swin Transformer at the ISSC encoder.
The multi-scale semantic features are first aggregated and then sent to the receiver through a wireless channel.
At the ISSC decoder, the received semantic features are decoded to reconstruct image segmentation, which reduces the transmission data and achieves reliable transmission.
\item We evaluate the ISSC system performance through simulation experiments. 
Simulation results demonstrate that the proposed ISSC system can provide robust transmission performance against channel variation, particularly at low SNR, and can reconstruct the image segmentation accurately with a high compression ratio.
Compared to the traditional methods, the proposed ISSC system can improve mean Intersection over Union~(mIoU) by 75\%.
\end{itemize}

The rest of this paper is structured as follows.
In Section II, the system model and problem formulation are presented.
We provide a thorough explanation of the proposed ISSC encoder and decoder in Section III.
In Section IV, the simulation results are shown and discussed, and
Section V concludes the paper. 

\section{SYSTEM MODEL}
We consider a single-link vehicle-to-vehicle image communication scenario where a front vehicle needs to transmit visual data to a rear vehicle that deploy an ISSC system over the IoV, as shown in Fig.~\ref{Traffic Scene}.
Due to the blocking of the front vehicle, the rear vehicle can not see the obstacle.
In this scenario, the front vehicle extracts and aggregates the semantic features of the image taken by the camera and sent to the rear vehicle.
Image segmentation has the category and location information of objects, so that the rear vehicle can reconstruct the image segmentation to make driving decisions.

\begin{figure}[t]
\centering
\includegraphics[width=9cm,height=2.6cm]{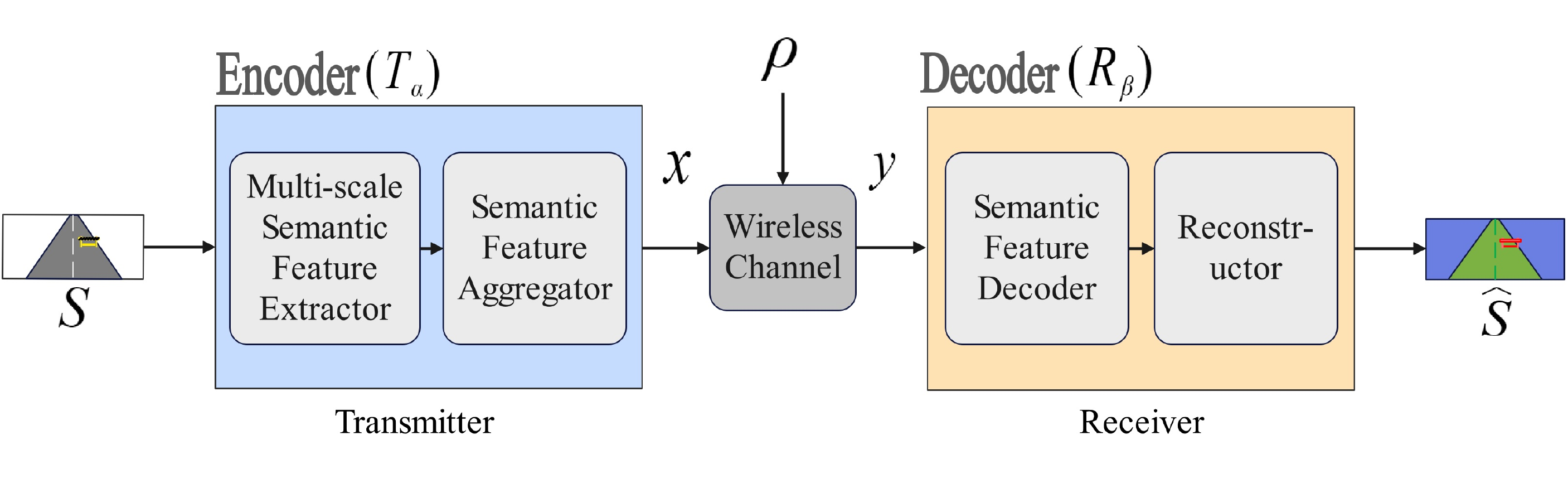}
\caption{The framework of the proposed ISSC system.}
\label{Simplify Structure}
\vspace{-0.55cm}
\end{figure}

The above ISSC system can be simplified as shown in Fig.~\ref{Simplify Structure}. 
In the system, the transmitter~(front vehicle) sends image semantic features to the receiver (rear vehicle) through a wireless channel.
The system consists of two components: (\textit{i}) the ISSC encoder at the transmitter side, which extracts semantic features from the input image for transmission; (\textit{ii}) the ISSC decoder at the receiver side, which decodes the received semantic features and utilize them to reconstruct the image segmentation.

\subsection{ISSC Encoder}
The ISSC encoder extracts and aggregates the semantic features from the input image  $\boldsymbol{S}\in \mathbb{R}^{H \times W \times 3}$ through a multi-scale semantic feature extractor and a semantic feature aggregator, where \textit{H}, \textit{W}, and 3 are the image width, the image height, and the number of  channels.
The input image is RGB format and every value range is in $[0, 255 ]$ which needs to be normalized to $[0, 1]$ by the normalization layer.
To extract the semantic features from the shallow to the deep, the processed image is put into to a multi-scale semantic feature extractor made up of neural networks. 
Then the semantic feature aggregator fuses the multi-scale semantic features and send them.

\begin{figure*}[htbp]
\centering{\includegraphics[width=18cm,height=6cm]{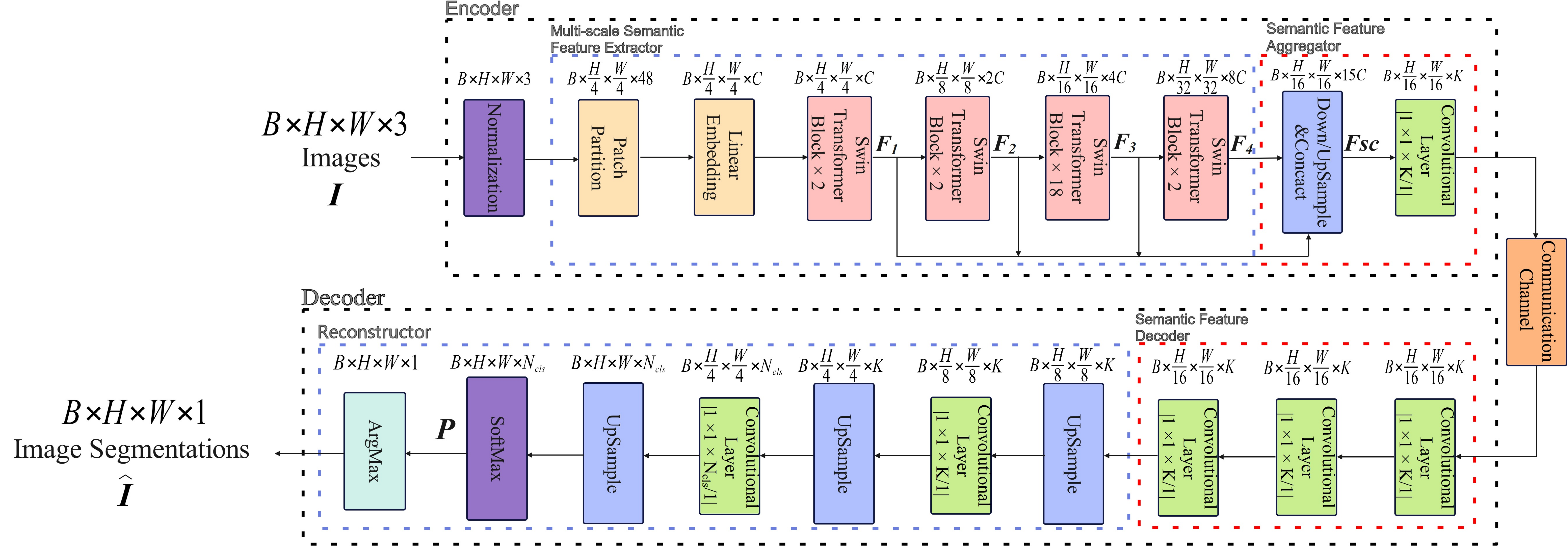}}
\caption{The architecture of the proposed ISSC system which consists of an endoder and a decoder.}
\label{network}
\vspace{-0.35cm}
\end{figure*}

We simplify the ISSC encoder parameter as $\boldsymbol{\alpha}$, thus, the relationship between the transmitted semantic features $\boldsymbol{x}$ and the input image $\boldsymbol{S}$ can be given by
\begin{equation}
\boldsymbol{x} = \boldsymbol{{T_\alpha }}(\boldsymbol{S}),\label{eq1}
\end{equation}
where $\boldsymbol{{T_\alpha }}(\boldsymbol{\cdot})$ indicates the function of the ISSC encoder.

When being transmitted over a wireless channel, the encoded semantic features will suffer channel fading and noise.
We consider the condition of a single communication link, where the received semantic features $\boldsymbol{y}$ at the ISSC decoder can be characterized as
\begin{equation}
\boldsymbol{y} =\boldsymbol{h} \cdot \boldsymbol{x} + \boldsymbol{\rho},\label{eq2}
\end{equation}
where $\boldsymbol{h}$ is the channel covariance coefficient, and $\boldsymbol{\rho} \sim \mathcal{N}\left(0, \sigma^2 \textbf{I} \right)$ is the Gaussian channel noise with variance $\sigma^2$ and $\textbf{I}$ is the identity matrix.

\subsection{ISSC Decoder}
The ISSC decoder is used to decode the received semantic features $\boldsymbol{y}$ and  reconstruct the image segmentation  $\boldsymbol{\widehat{S}}$, consists of a semantic feature decoder and a reconstructor. 
First, the semantic feature decoder needs to reduce the influence of channel noise.
Then, the reconstructor needs to get the image segmentation $\boldsymbol{\widehat{S}}$ with the same length and width of the input image. $\boldsymbol{\widehat{S}}$  classifies the pixels in $\boldsymbol{S}$ into ${N_{cls}}$ categories where ${N_{cls}}$ is the number of object categories concerned in $\boldsymbol{S}$ (pedestrians, vehicles, trucks, etc.).
We simplify the ISSC decoder parameter as $\boldsymbol{\beta}$, similar to the ISSC encoder.
At this point, the association between the output image segmentation $\widehat{S}$ and the received semantic features y can be given by 
\begin{equation}
\boldsymbol{\widehat{S}}=\boldsymbol{R_{\beta}}(\boldsymbol{y}),\label{eq3}
\end{equation}
where $ \boldsymbol{{R_{\beta}}}(\boldsymbol{\cdot} )$ indicates the function of the ISSC decoder.

\subsection{ISSC Objective}
The objective of the ISSC system is to reconstruct the image segmentation of the vehicle precieved image at the receiver. 
Since our system focuses on semantic-level recovery, we use the mIoU metric rather than the bit error rate in traditional communications to evaluate the system performance, which is given by
\begin{equation}
{\rm mIoU}=\frac{1}{N_{cls}} \sum_{i=1}^{N_{cls}} \frac{P \bigcap G}{P \bigcup G},\label{eq4}
\end{equation}
where \textit{P} is the set of pixel regions predicted by the ISSC decoder for a certain category object, and \textit{G} is the actual set of pixel regions for this category object.
The higher the mIoU, the better the system performance. 
\section{ISSC encoder and decoder design}
In the section, we will introduce the proposed ISSC system architecture, as show in Fig.~\ref{network}.
Firstly, to efficiently refine visual semantics at various scales, ISSC encoder at the transmitter uses a multi-scale semantic feature extractor and a semantic feature aggregator.
At the receiver, the ISSC decoder employs a semantic feacture decoder and a reconstructor to perform semantic feature decoding and image segmentation reconstructing, respectively.

\subsection{Multi-scale Semantic Feature Extractor and Semantic Feature Aggregator}
The input of the ISSC system is expressed as $\boldsymbol{I}\in \mathbb{R}^{B \times H \times W \times 3}$, denoted a batch of images with \textit{B} being the batch size.
Before entering the ISSC encoder, the pixel values of the images are normalized  to [0, 1] to hasten model convergence and decrease the computing cost. 
We initially divide every $H{\times}W{\times}3$-sized image into $4\times4$-sized patches, because the dense prediction task at the receiver benefits from using smaller patches.
Following a linear embedding layer, the patch channel can be converted into any number \textit{C}. 

Considering that most of the previous works \cite{bourtsoulatze2019deep,kurka2020deepjscc,yang2022deep,yang2021semantic,kang2022task,hu2022robust} on image semantic communication systems are based on convolutional neural networks (CNNs) whose ability to extract semantics is limited by the size of the convolution kernel\cite{dosovitskiy2020image}.
As a substitute for CNNs,  Swin Transformer\cite{liu2021swin} has shown strong feature extraction ability and low computational complexity.
Therefore, we adopt Swin Transformer in the multi-scale semantic feature extractor to extract image semantic features from shallow to deep layers as shown in Fig.~\ref{network}.
Swin Transformer contains four stages and each has even number of Swin Transformer blocks (STBs). 
STB utilizes multi-head self-attention (MSA) mechanism to obtain larger receptive fields than CNNs, and the STB architecture can be shown in Fig.~\ref{STB}.
Furthermore, to reduce computation overhead, STB divides the input into many independent small regions, called as window partition as shown in Fig.~\ref{window} left, and computes the self-attention by window based MSA (W-MSA) module  separately.
Also, to enable the interaction of adjacent pixels in different windows, each regular STB is followed a STB with shifted window based MSA (SW-MSA) module  which shifting the windows by $(\left\lfloor {\frac{M}{2}} \right\rfloor ,\left\lfloor {\frac{M}{2}} \right\rfloor )$ pixels versus the proceeding STB as illustrated in Figs.~\ref{STB}, and~\ref{window}.

Given an input $ \boldsymbol{Z} \in {\mathbb{R}^{H \times W \times C}}$, STB first 
partitions $ \boldsymbol{Z}$ into  $ \frac{{HW}}{{{M^2}}}$ non-overlapping $M {\times} M$-sized windows. 
Then, STB computes the self-attention of pixels separately in each window. For a window feature  $ \boldsymbol{Z'} \in {\mathbb{R}^{{M^2} \times C}}$,  $\boldsymbol{Q}$,  $\boldsymbol{K}$, and  $\boldsymbol{V}$ represent \textit{query}, \textit{key}, and \textit{value}, respectively, and can be given by
        \begin{equation}
                   \boldsymbol{Q} = \boldsymbol{Z'{W_{^Q}}},  \boldsymbol{K = Z'{W_{^K}}}, \boldsymbol{V = Z'{W_{^V}}},\label{qkv}
        \end{equation}
where $\boldsymbol{W_{^Q}}$, $\boldsymbol{W_{^K}}$ and $\boldsymbol{W_{^V}}$ are projection matrices shared by different windows.
Given $\boldsymbol{Q}$, $\boldsymbol{K}$,  $\boldsymbol{V} \in {\mathbb{R}^{{M^2} \times d}}$, with \textit{d} being the 
\textit{query/key} dimension, the self-attention output matrix of a window can be computed by
    \begin{equation}
    \operatorname{Attention}(\boldsymbol{Q},\boldsymbol{K},\boldsymbol{V}) = \operatorname {SoftMax}(\boldsymbol{QK}^{T}/\sqrt{d}+\boldsymbol{L}) \boldsymbol{V},\label{attention}
    \end{equation}
where  $ \boldsymbol{L} \in {\mathbb{R}^{M^2\times M^2}}$ is the learnable relative position bias. In practice, STB uses multiple heads for computing self-attention in parallel, and then concatenates the results as the MSA output. 
\begin{center} 
\begin{figure}[tbp]  
\centering
	\subfigure[Two successive STBs]{
		\includegraphics[width=5cm,height=4cm]{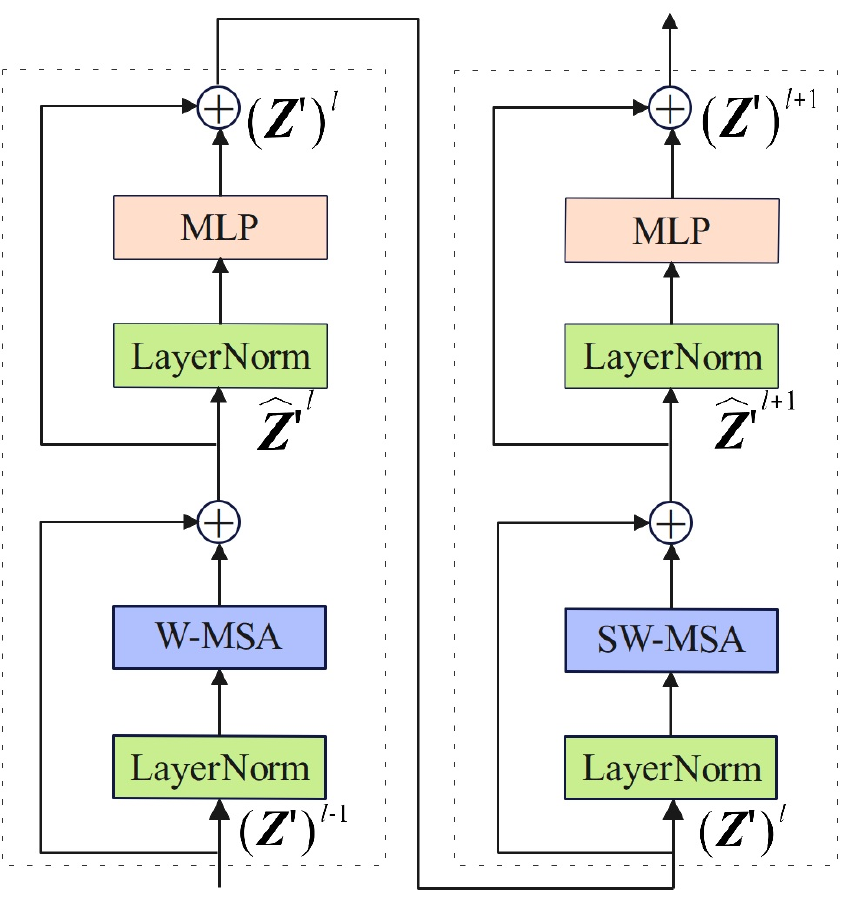}
		 \label{STB}
		 }
			\quad
	\centering
	\subfigure[Window partition and shifted window partition]{
		\includegraphics[width=5.5cm,height=3.5cm]{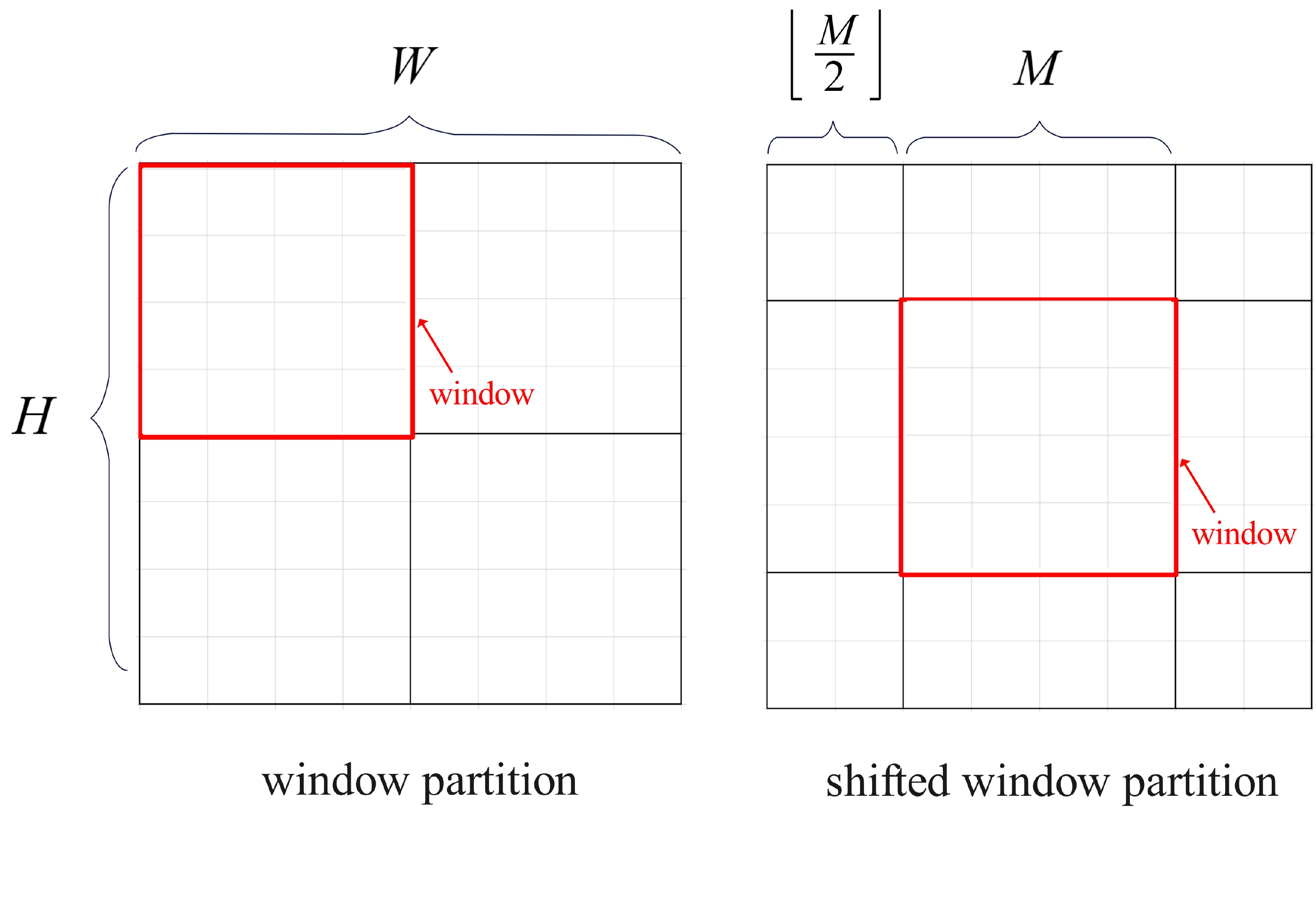}
		\label{window} 
	} 
	\quad
	\centering
	\subfigure[MLP]{
		\includegraphics[width=2cm,height=4cm]{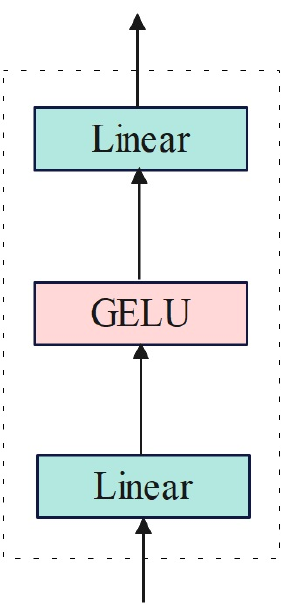}
		\label{MLP} 
	}
	
	\caption{(a) The architecture of two successive STBs; (b) Illustration of the window partition and shifted window partition where \textit{H} = 8, \textit{W} = 8, and \textit{M} = 4; (c) The architecture of the MLP.}
	\label{STB&MLP}
\vspace{-0.35cm}
\end{figure}
\vspace{-0.8cm}
\end{center}
 
STB also includes a multi-layer perceptron (MLP) layer , as shown in Fig.~\ref{MLP}, and two LayerNorm (LN) layers with the residual connection. The whole process of  $ \boldsymbol{Z'}$ passes through two successive STBs are computed as
     \begin{equation}   
    \begin{aligned}
         &\boldsymbol{\widehat{Z'}}^{l}={\rm W\mbox{-}MSA(LN}((\boldsymbol{Z'})^{l-1})) + (\boldsymbol{Z'})^{l-1}, \\       
         &(\boldsymbol{Z'})^{l}={\rm MLP(LN}(\boldsymbol{{\widehat{Z'}}}^{l})) + \boldsymbol{{\widehat{Z'}}}^{l},   \\ \label{process1}
      &\boldsymbol{\widehat{Z'}}^{l+1}={\rm SW\mbox{-}MSA(LN}((\boldsymbol{Z'})^{l})) + (\boldsymbol{Z'})^{l}, \\       
         &(\boldsymbol{Z'})^{l+1}={\rm MLP(LN}(\boldsymbol{{\widehat{Z'}}}^{l+1})) + \boldsymbol{{\widehat{Z'}}}^{l+1},  
    \end{aligned}
    \end{equation}
where ${l}$ is the layer number of STB.


In the multi-scale semantic feature extractor, we get four stages' features as shown in Fig.~\ref{network}, which are denoted by  $\boldsymbol{F_1}$,   $\boldsymbol{F_2}$,  $\boldsymbol{F_3}$, and  $\boldsymbol{F_4}$. 
From $\boldsymbol{F_1}$ to $\boldsymbol{F_4}$, the features become increasingly rich in image semantics.

In the semantic feature aggregator, the ISSC samples  $\boldsymbol{F_1}$,  $\boldsymbol{F_2}$, and  $\boldsymbol{F_4}$ to the size of $\boldsymbol{F_3}$ and concatenates them to obtain the features $\boldsymbol{F_{sc}}$.
Then, the features $\boldsymbol{F_{sc}}$ pass a convolutional layer with \textit{K} filters of size 1 and stride 1, the value of \textit{K} is used to adjust the ISSC encoder compression ratio. 

\subsection{Semantic Feature Decoder and Reconstructor}
At the receiver, the semantic feature decoder consists of three convolutional layers as shown in Fig.~\ref{network}, which are used to reduce the impact of noise. 
Each of the convolutional layers has  ${K}$ filters  of size 1 and stride 1. 

The reconstructor consists of two convolutional layers, three upsample layers,  a softmax activation, and a argmax layer.
The first  convolutional layer  has  ${K}$ filters  of size 1 and stride 1, and the second convolutional layer  has  ${N_{cls}}$ filters  of size 1 and stride 1.
Three upsample layers interspersed between two convolutional layers gradually make the length and width of the features the same as the input.
By passing a softmax activation, we get the features  $ \boldsymbol{P} \in {\mathbb{R}^{B \times H \times W \times N_{cls}}}$, the value of the last dimension is the predicted probability for pixel multi-category classification. 
Finally, we preform the argmax operatioin to get the image segmentations  $\boldsymbol{\widehat I}\in \mathbb{R}^{B \times H \times W \times 1}$ and the value of the last dimension is the category subscript with the maximum predicted probability value.

The goal of the reconstructor is to categorize the image pixels correctly, therefore, we utilize the cross entropy of multi-category classification for each pixel as the loss function, which can be given by
\begin{equation}
{\mathcal L}_{CE} =\sum\limits_{i = 1}^{{N_{cls}}} {({y_{i}}\log ({p_{i}}))} ,
\end{equation}
where 
$p_i$ is the probability that the pixel is classified as category $i$ and $y_i$ is the classification indicator with value 0 or 1.
For a batch of images, the loss function of the whole ISSC system can be given by 
\begin{equation}
{\mathcal L}_{ISSC} = \frac{{ - \sum\limits_{n = 1}^B {\sum\limits_{m = 1}^{H \times W} {\mathcal L}_{CE} }} }{{B \times H \times W}}.\label{eq6}
\end{equation}
Given the loss function, the ISSC system is trained by an end-to-end method. 
Thus, the semantic feature decoder can recede the channel affect when decoding the received semantic features.



\section{SIMULATION AND PERFORMANCE ANALYSIS}
To evaluate the performance of the proposed ISSC system, we use the urban traffic scenario dataset Cityscaspes\cite{cordts2016cityscapes} for training and testing.
The Cityscapes dataset contains 2975 training images and 1525 test images, and every image is in RGB format with $2048\times 1024$ pixels.
We adopt random cropping, random flipping, and optical distortion for training images to achieve data augmentation.
We train the model under additive white Gaussian noise channel~(AWGN) with the SNR varies randomly between 1 and 20 dB and Adam algorithm is employed to optimize the model. 
Additionally, we use an online hard sample mining technique that only trains pixels with confidence bellow 0.7 and keeps training at least 100,000 pixels.
TABLE \uppercase\expandafter{\romannumeral1} lists the simulation parameters. 

For the purpose of comparison, we implement the traditional modular wireless transmission mode, which adopts JPEG and PNG for source coding, 2/3-rate Low-Density Parity-Check Codes (LDPC) for channel coding, and 4-QAM, 16-QAM, and 64-QAM for modulation.
At the receiver, we employ the DeepLabv3+ and the OCRnet to get the received image segmentations.
For a fair comparison, we set the compression ratio $r=3$ for both ISSC system and traditional methods.
\vspace{-0.35cm}
\begin{table}[tbp!]
  \centering
  \caption{SIMULATION PARAMETERS}
  \resizebox{\linewidth}{3cm}{
    \begin{tabular}{|c|c|c|c|}
    \hline
          \textbf{Module} & \textbf{Layer Name} & \textbf{Parameter} & \textbf{Values} \\
    \hline
     \multirow{7}{*}{\makecell[c]{Multi-scale \\ Semantic Feature\\ Extractor}} & \multirow{7}{*}{\makecell[c]{4${\times}$Swin Transformer\\ Stage}} & depths & 2, 2, 18, 2 \\
\cline{3-4}                 &       & head number & 3, 6, 12, 24 \\
\cline{3-4}                 &       & \makecell[c]{embedding\\ dimension} & 96 \\
\cline{3-4}                 &       & window size & 7 \\
\cline{3-4}                 &       & patch size & 4 \\
\cline{3-4}                 &       & MLP ratio & 4 \\
\cline{3-4}                 &       & activation & GELU \\
\hline          \multirow{4}{*}{\makecell[c]{\makecell[c]{Semantic Feature\\ Aggregator}}} & 4${\times}$Down/UpSample & sampling rate & 1/4, 1/2, 1, 2 \\
\cline{2-4}                 & \multirow{3}{*}{Convolutional Layer} & kernel size & 1${\times}$1 \\
\cline{3-4}                 &       & stride & 1 \\
\cline{3-4}                 &       & number & K \\
    \hline
    AWGN  & AWGN  & $\rm{SNR_{train}}$ (dB) & 1-20 \\
    \hline
    \multirow{3}{*}{Semantic Feature Decoder} & \multirow{3}{*}{3${\times}$Convolutional Layer} & kernel size & 1${\times}$1 \\
\cline{3-4}                 &       & stride & 1 \\
\cline{3-4}                 &       & number & K, K, K \\ 
\hline           

 \multirow{4}{*}{\makecell[c]{\makecell[c]{Reconstructor}}} &  3${\times}$UpSample & sampling rate &  2, 2, 4 \\
\cline{2-4}                 & \multirow{3}{*}{2${\times}$Convolutional Layer} & kernel size & 1${\times}$1 \\
\cline{3-4}                 &       & stride & 1 \\
\cline{3-4}                 &       & number & K, $\rm{N_{cls}}$ \\

    \hline
    \end{tabular}}%
  \label{tab:addlabel}%
\vspace{-0.35cm}
\end{table}%
\vspace{0.3cm}
\begin{figure}[tbp]
\centering
\subfigure[Image segmentation visual display of the traditional method using JPEG and the OCRnet]{
\includegraphics[scale=0.14]{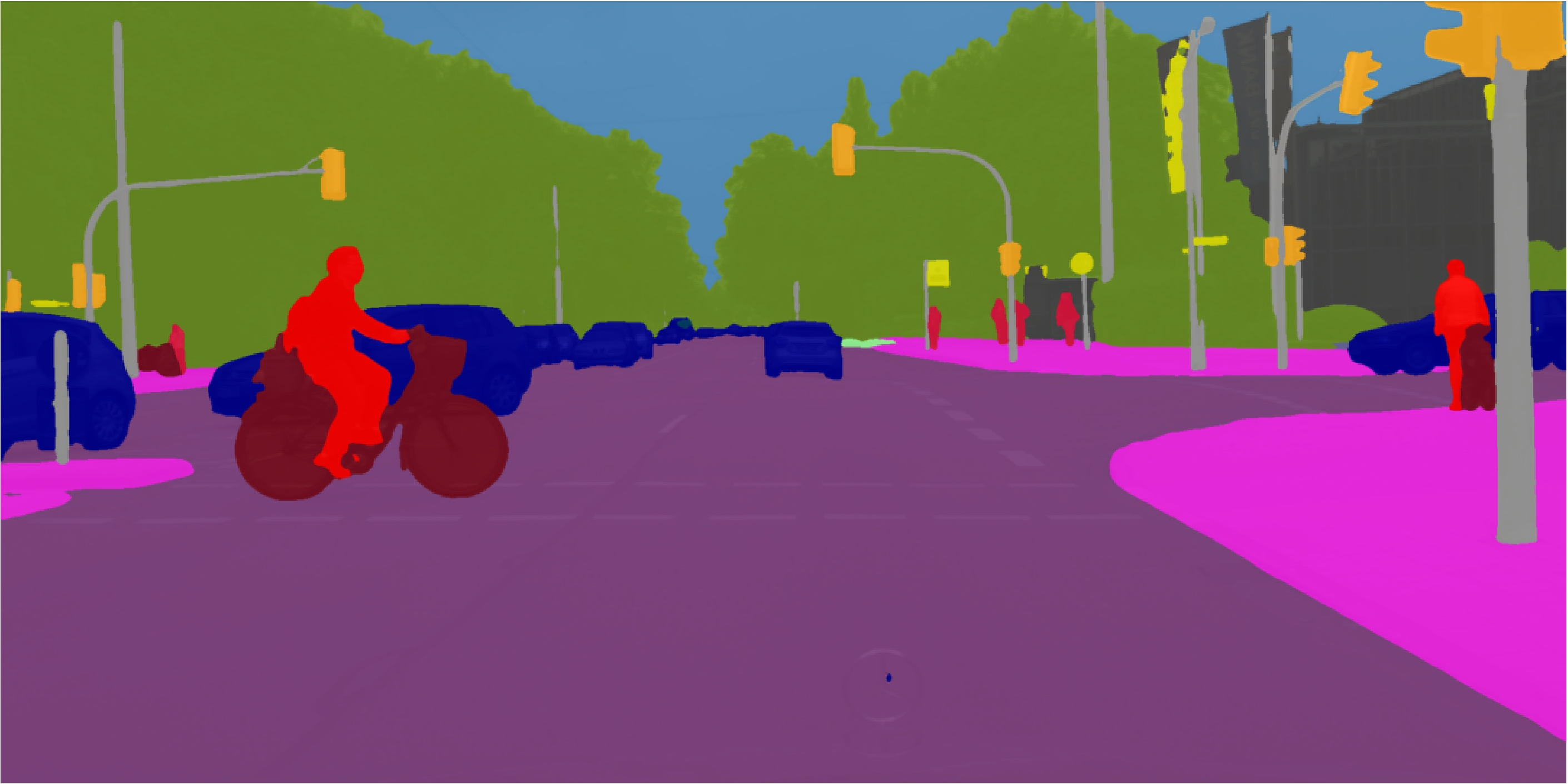} \label{16QAM_23LDPC_JPEG rate=3}
}
\quad
\subfigure[Image segmentation visual display of the ISSC system ]{
\includegraphics[scale=0.14]{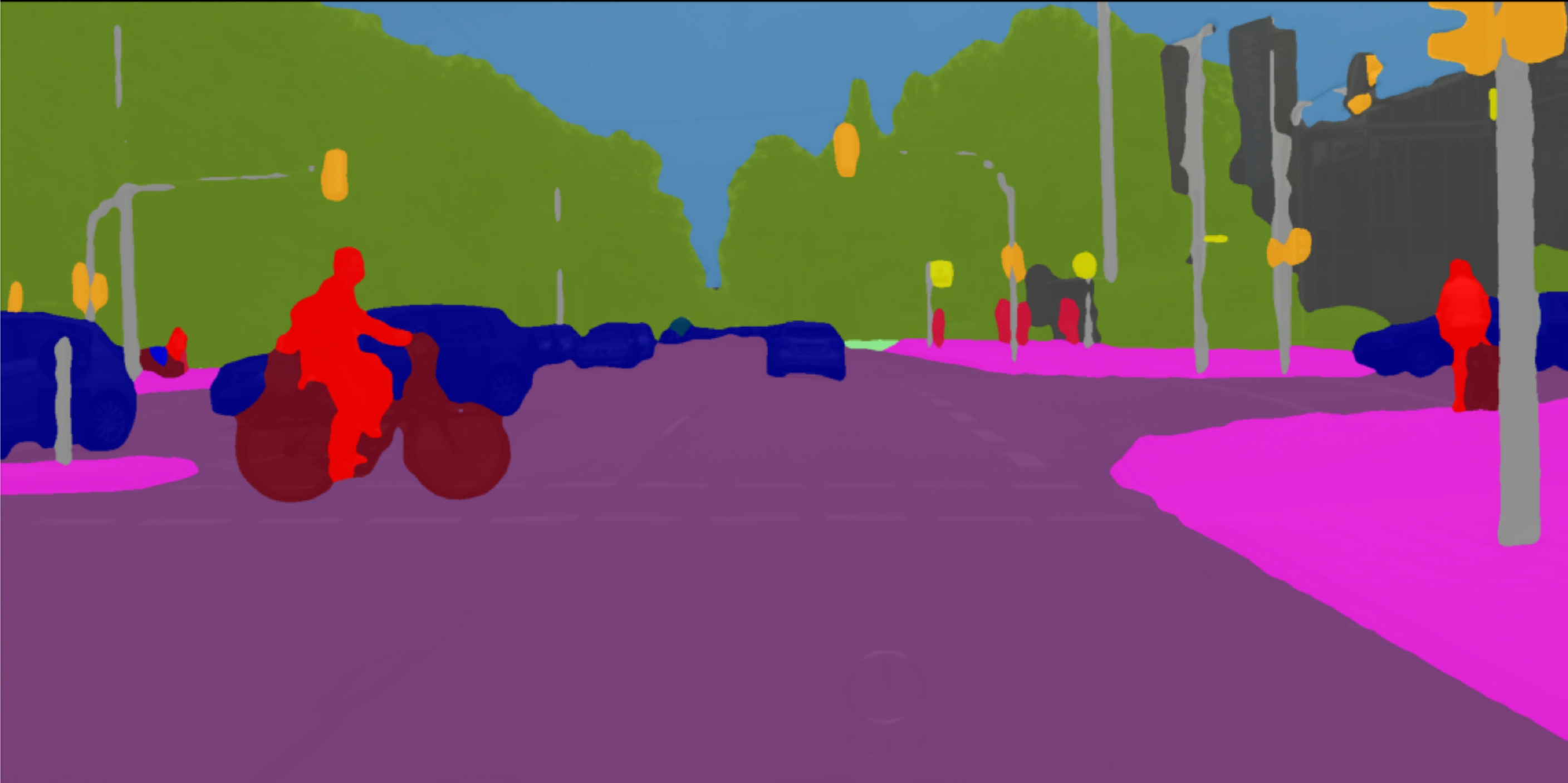} \label{ratio=3_16dB_ISSC}
}
\caption{Image segmentation visual comparison between the traditional method and the ISSC system under an AWGN channel at $r=3$.}
\label{Visual comparison}
\vspace{-0.65cm}
\end{figure}

Fig.~\ref{Visual comparison} shows the image segmentation comparison of the traditional method using 16-QAM modulation and the ISSC system under an AWGN channel with 22 dB SNR.
In this case, ISSC system only transmits 1/3 less data than the traditional method using 2/3 rate LDPC.
From Fig.~\ref{Visual comparison}, we see that the traditional method has a better segmentation at details than ISSC system. This is because that the traditional method can completely reconstruct the low compression ratio coded image under high SNR.
We can also see that the segmentation result of the ISSC system for the vital objects (vehicle, human, road, etc.) are closed to those of the traditional method, which demonstrates the advantage of the proposed ISSC system.

\begin{figure}[tbp]
\centering{\includegraphics[width=9cm,height=6cm]{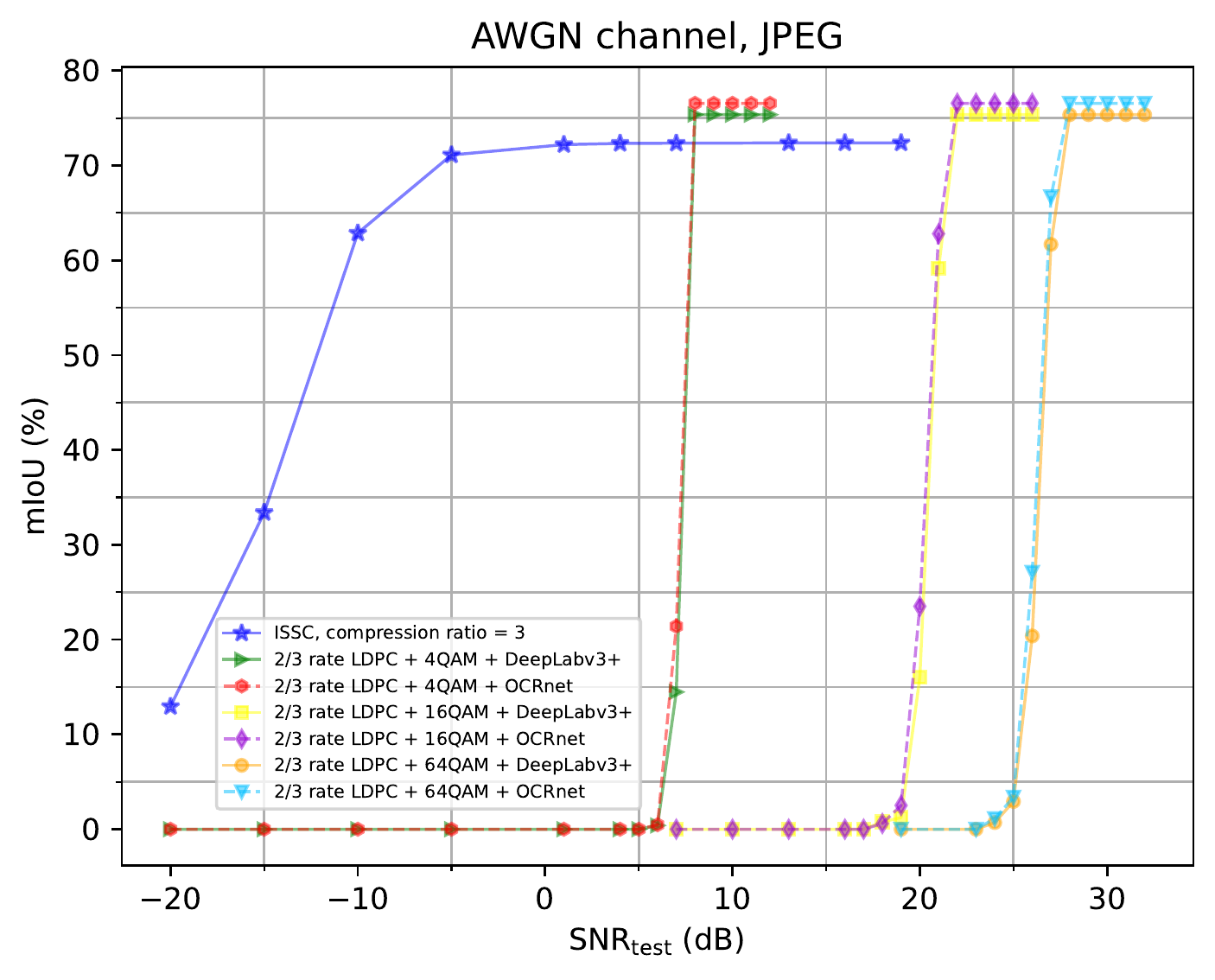}}
\caption{mIoU comparison of the ISSC system with the traditional methods using JPEG.}
\vspace{-0.35cm}
\label{mIoU_JPEG}
\end{figure}

\begin{figure}[tbp]
\centering
\subfigure[18 dB]{
\includegraphics[scale=0.11]{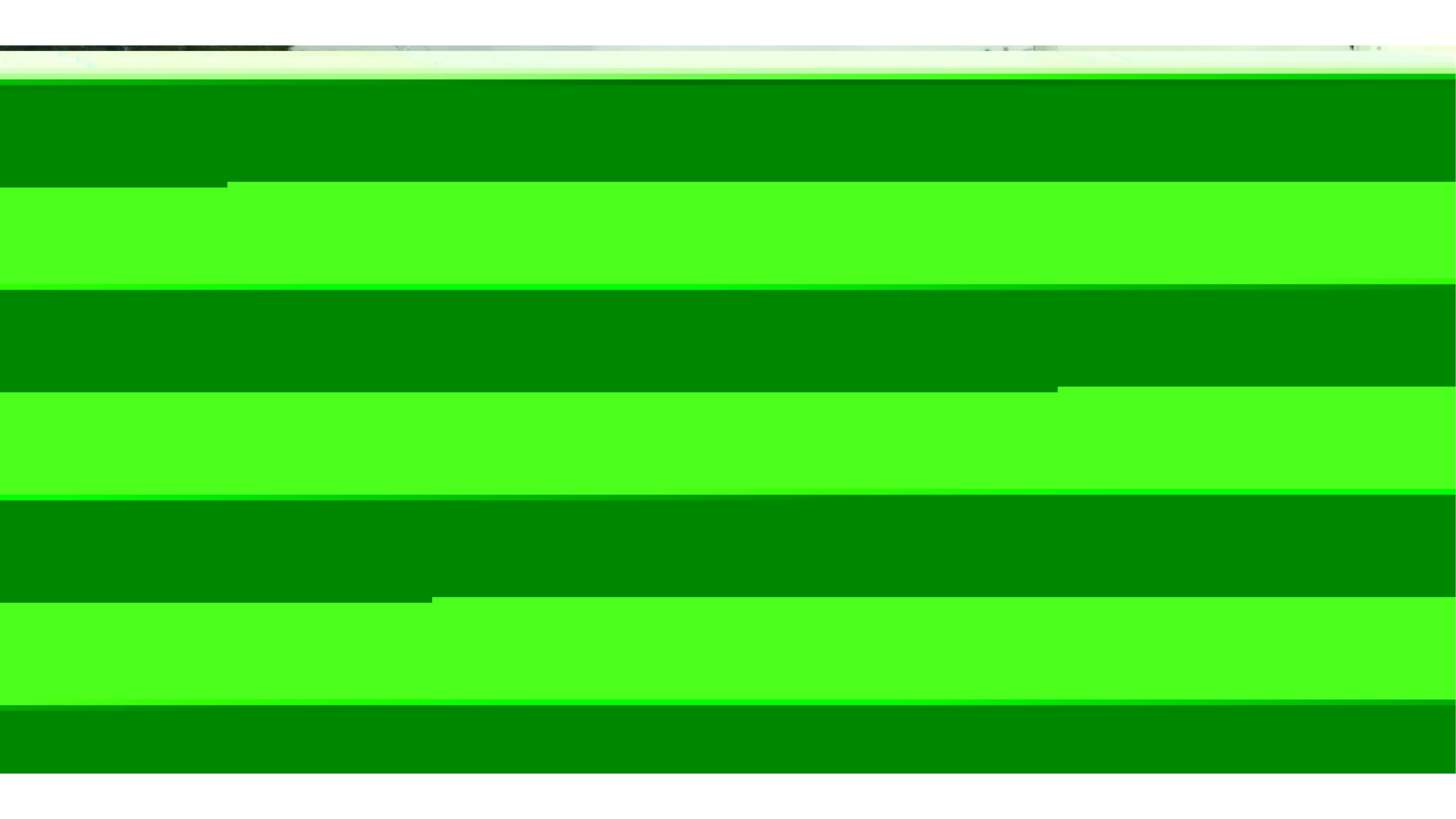} \label{18dB}
}
\quad
\subfigure[19 dB]{
\includegraphics[scale=0.11]{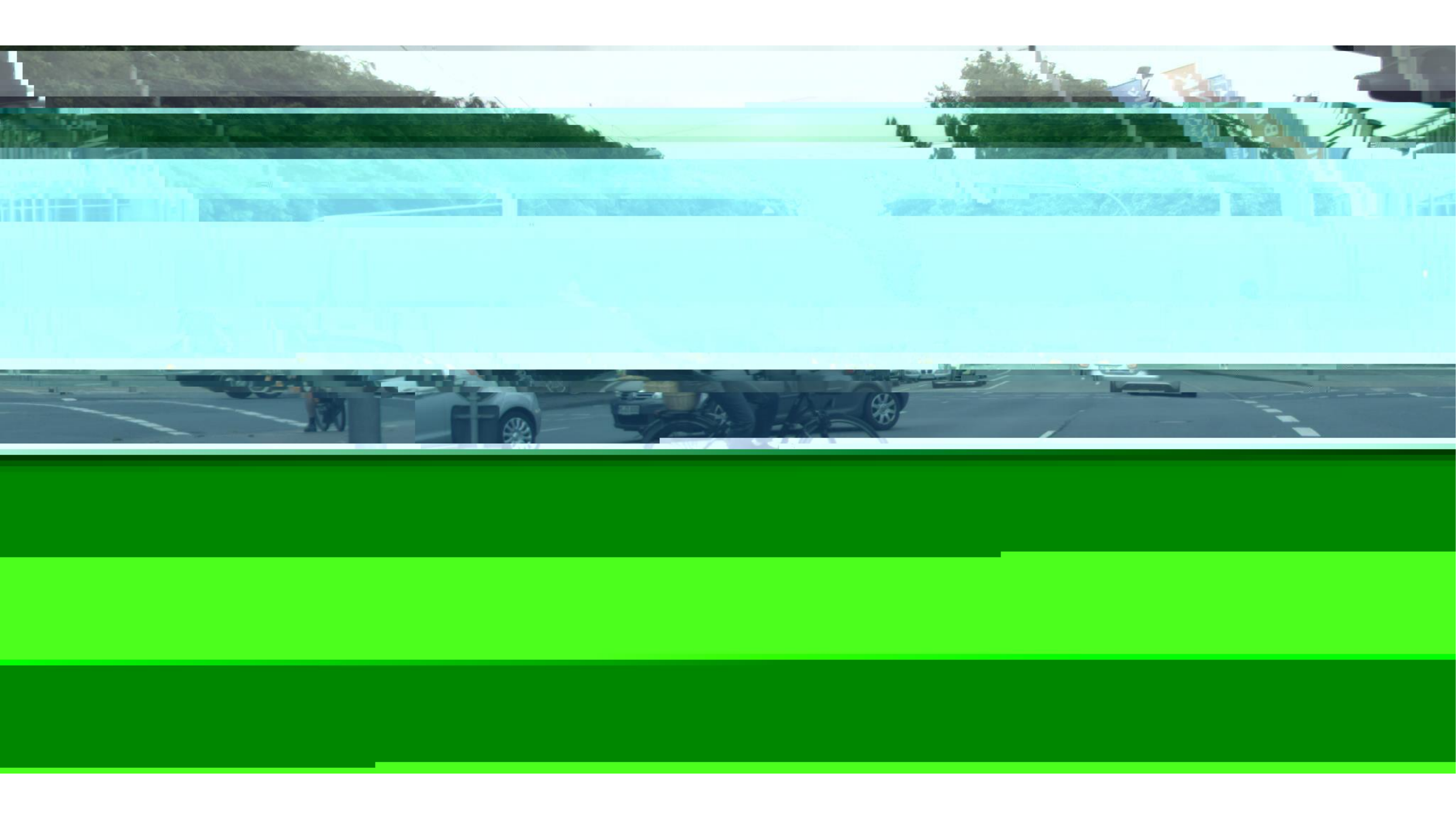} \label{19dB} 
}
\quad
\subfigure[20 dB]{
\includegraphics[scale=0.11]{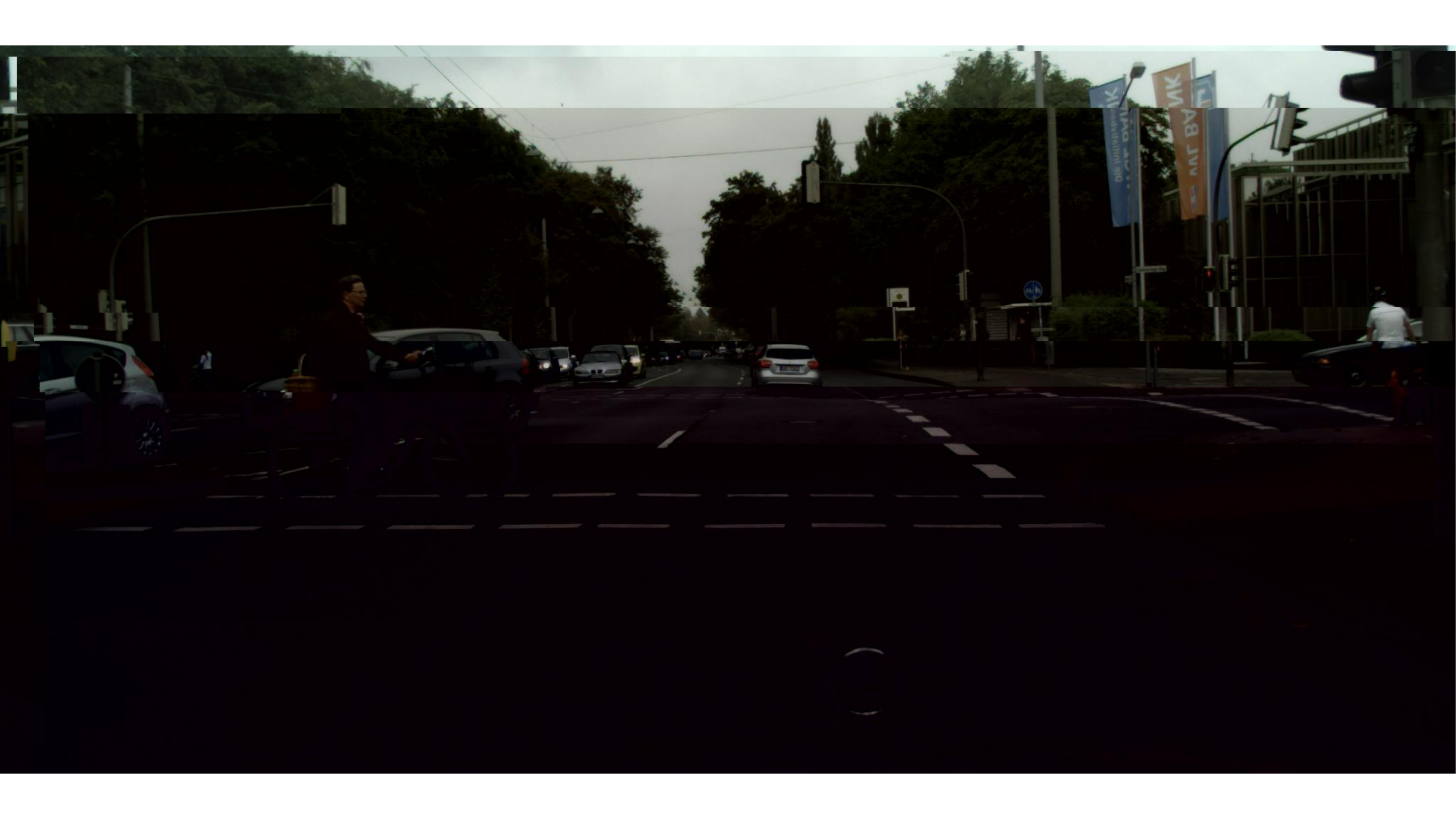} \label{20dB} 
}
\quad
\subfigure[21 dB]{
\includegraphics[scale=0.11]{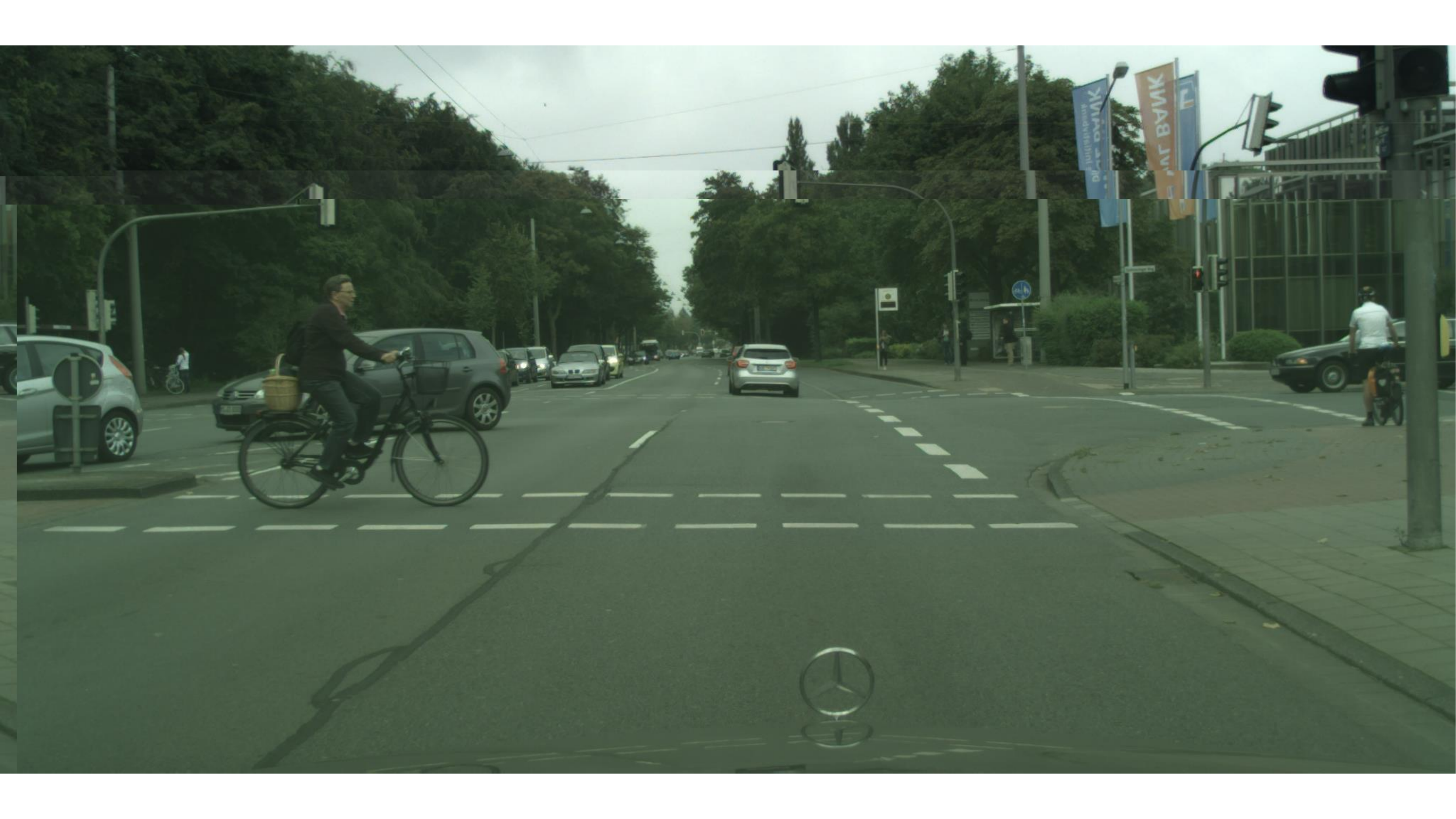} \label{21dB} 
}
\caption{The ``cliff effect'' of the traditional methods over the AWGN channel.}
\label{cliff effect}
\vspace{-0.55cm}
\end{figure}

In Fig.~\ref{mIoU_JPEG}, we display the mIoU comparsion with different SNR values.
We can observe that, as the channel SNR increases, the mIoU of both the ISSC system and the traditional methods increase due to the gentler noise.
We can also see that the mIoU of the ISSC system is more robust with the steadier mIoU, in contrast to the traditional methods sharply drop when the channel quality deteriorates. 
The reason for the steep line of the traditional methods is that, JPEG encodes the image into flag and data bits, which would be transmitted incorrectly at low SNR. 
And the wrong flag bits will severely impact the image reconstruction and damage the image segmentation.
This phenomenon is known as ``cliff effect''\cite{bourtsoulatze2019deep}. 
From Fig.~\ref{mIoU_JPEG} we see that, ``cliff effect'' is avoided by ISSC system with transmitting semantic features. 
This is because the ISSC system can effectively extract image semantic features, thus achieving more robust transmission against channel variation. \color{black}
Fig.~\ref{cliff effect} shows the visual display of the ``cliff effect'' in a traditional method that uses JPEG and 16-QAM under the AWGN channel with 18 dB, 19 dB, 20 dB, and 21 dB SNR.
Figs.~\ref{Visual comparison}, \ref{mIoU_JPEG}, and~\ref{cliff effect} show that the proposed ISSC system can effectively avoid the ``cliff effect''.
\color{black}

\begin{figure}[tbp]
\centering
\includegraphics[width=9cm,height=6cm]{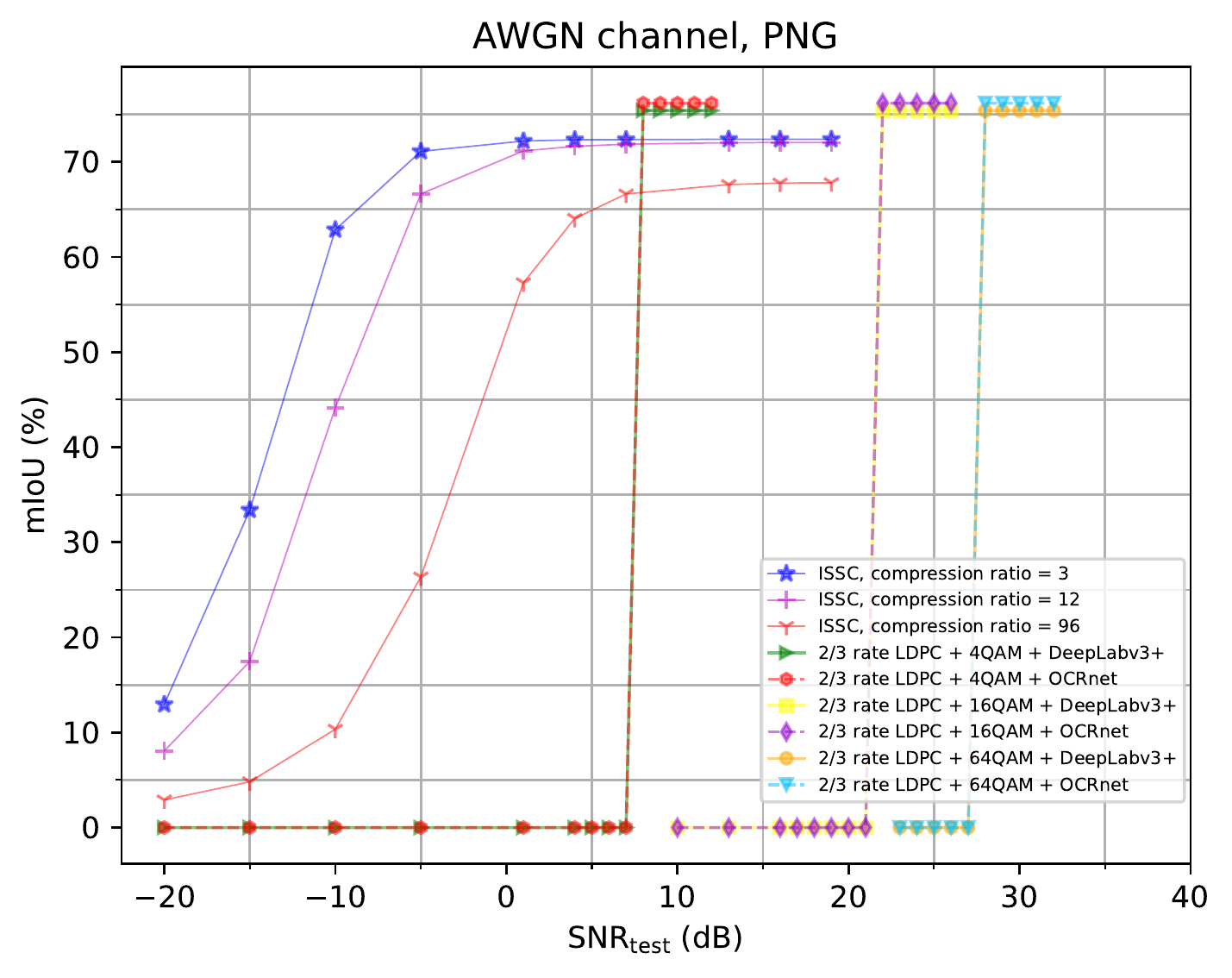}
\caption{mIoU comparison of the ISSC system with the traditional methods using PNG.}
\label{mIoU_PNG}
\vspace{-0.45cm}
\end{figure}

\begin{figure}[tbp]
\centering
\includegraphics[width=9cm,height=6cm]{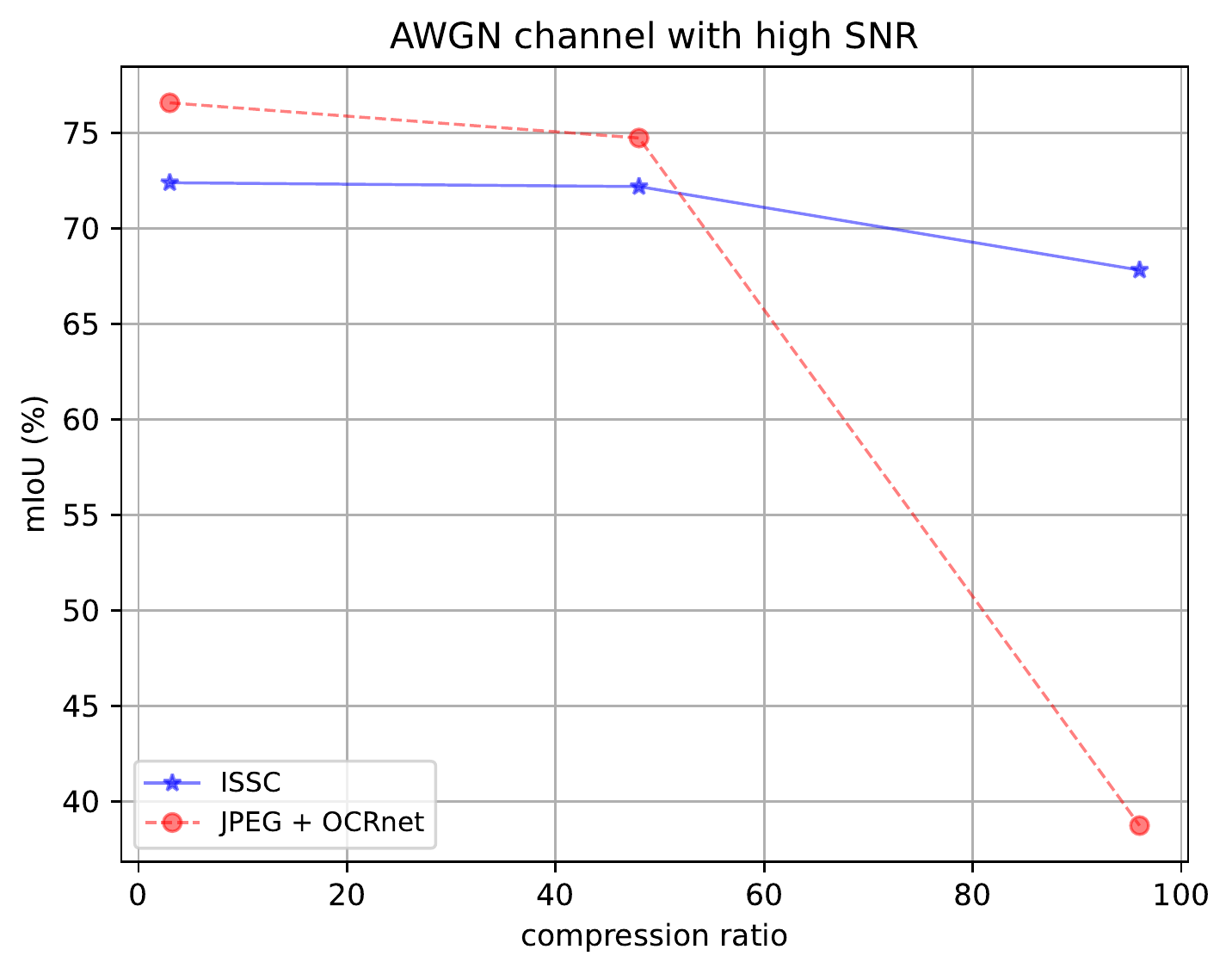}
\caption{Effect of compression ratio variation on the performance of the ISSC system and the tradition method using JPEG.}
\label{commpression ratio}
\vspace{-0.55cm}
\end{figure}

In Fig.~\ref{mIoU_PNG}, we further implement PNG as source coding in traditional methods.
From Fig.~\ref{mIoU_PNG}, we can see that, using PNG gets the similar curves as that of JPEG but the steep lines have no intact outputs because the damaged PNG image cannot be segmented. \color{black}
From Fig.~\ref{mIoU_JPEG} and Fig.~\ref{mIoU_PNG}, we can see that the ISSC system has a better performance at low SNR, because ISSC system can learn the channel variation during the training stage and obtain the error correction ability at the semantic-level. 
In addition, the traditional methods performs better than ISSC system at high SNR, mainly on the edge of the objects.
The reason is that, at low compression ratios, the traditional methods can reconstruct the nearly intact image which provides more semantic information during the image segmentation stage.
In addition, we can see that ISSC system has similar curves with the change of SNR under different compression ratios, 
and for a given SNR, the performance increases with the decrease of compression ratio. 
This is due to the fact that, ISSC system  with different compression ratios uses the same structure as the multi-scale semantic feature extractor. As the aggregator uses the convolutional layers to retain more semantics features, the receiver can reconstruct the image segmentation better.

Fig.~\ref{commpression ratio} illustrates how the performance of the ISSC system and the traditional method varies as compression ratio increases.
From Fig.~\ref{commpression ratio} we see that, as the compression ratio increases, the mIoU of both the ISSC system and the traditional method decrease, but at high compression ratios, ISSC system maintains an acceptable level in terms of mIoU rather than the traditional method.
Fig.~\ref{commpression ratio} shows that the proposed ISSC system can achieve a higher compression ratio, thus reducing the transmission data amount.



\section{CONCLUSION}
In this paper, we propose an ISSC system for vehicular wireless image transmission, which transmits the semantic features of images to achieve more reliable and efficient communication than traditional methods. 
Additionally, the neural networks based on Swin Transformer which broadens the receptive area of image data and extracts semantic features more effectively.
In the ISSC system, the codecs are jointly designed and trained to achieve global optimization of the model parameters.
According to experiments, the ISSC system performs better in low SNR, and its performance does not suffer much from deteriorating channel conditions and high compression ratios.
It is a competitive choice for image transmission over the IoV since it can handle the complex  traffic communication environment and reduce the amount of data transmission.

\def\baselinestretch{0.80}
\bibliographystyle{IEEEtran}
\bibliography{IEEEabrv,ref1}
\end{document}